\documentclass[12pt]{article}
\usepackage{amssymb}
\usepackage{amsmath}
\begin{document}
\title{Anti-de-Sitter Island-Universes from 5D Standing Waves}
\author{{\bf Merab Gogberashvili$^{(1,2)}$ and Douglas Singleton$^{(3,4)}$}
\\ \\
{\small (1) Andronikashvili Institute of Physics,
6 Tamarashvili Street, Tbilisi 0177, Georgia} \\
{\small (2) Javakhishvili State University,
3 Chavchavadze Avenue, Tbilisi 0128, Georgia}\\
{\small (3) California State University, Fresno, Physics Department,
Fresno, CA 93740-8031, USA} \\
{\small (4) VNIIMS,
Peoples' Friendship University of Russia, Moscow 117198, Russia}} 
\maketitle
\begin{abstract}
We construct simple standing wave solutions in a 5D space-time with
a ghost-like scalar field. The nodes of these standing waves are
`islands' of 4D anti-de-Sitter space-time. In the case of
increasing (decreasing) warp factor there are a finite (infinite)
number of nodes and thus a finite (infinite) number of anti-de-Sitter
island-universes having different gravitational and cosmological
constants. This is similar to the landscape models, which postulate
a large number of universes with different parameters. 
\end{abstract}
\vskip 0.5cm


\section{Introduction}

The infinite extra dimension brane theories \cite{brane1, brane2, brane3, brane4} have been
one of the most active areas of study in the past decade. A concise
overview of these models can be found in \cite{Rub}. The original
brane models had 5D space-times with 4D $\delta$-function sources
which were positive or negative tension thin branes. Later works
replaced the $\delta$-function sources with localized energy
momentum tensors of finite thickness which were called thick branes.
Examples of thick brane models included fluid-like matter sources
\cite{GoSi, GoSi1}, scalar field matter sources \cite{Bronn, dzh-sing}, and phantom
scalar field sources \cite{Koley, Koley1, Koley2, Phantom, Phantom1, Phantom2}.
Most brane world models focused on static solutions, i.e. both
metric and any fields being time independent. An exception to this
are S-brane solutions \cite{S, S1, S2, S3, S4} which are used for cosmological
studies.

In this paper we present a simple standing gravitational wave
solution as a possible 5D brane model. The existence of this
solution requires a negative cosmological constant and  a
phantom/ghost-like scalar field (i.e. a field with a negative sign
in front of the field kinetic energy term of the Lagrangian) in the
bulk. Such ghost-like scalar fields are generally problematic since
they tend to make the system unstable. To counter this criticism we
will show that our model can be embedded in a 5-dimensional
Weyl model of gravity \cite{Weyl, Weyl1, Weyl2, Weyl3}. Because of the
geometrical origin of the coupling of the scalar to the space-time and 
since the Weyl scalar does not couple with other matter fields we are
able to avoid the usual problems of instability due to ghost fields.

We assume that the ghost-like field vanishes on
the brane at the origin. This condition leads to a quantization of
the oscillation frequency of the standing wave which in turn makes the
scalar field vanish at a finite or infinite number of nodes of the
standing wave depending on whether the warp factor of the metric is increasing or decreasing.
By taking the matter fields to bind to the standing waves nodes one has
a finite or infinite number of anti-de-Sitter island-universes, each
with different parameters. In the simple model presented here the
different parameters are the effective 4D gravitational constant and
the 4D cosmological constant. These two 4D constants are fixed in terms of the 5D constants
and the value of the warp factor at the nodes.

For the case of an increasing warp factor, and a finite number of anti-de-Sitter
`islands', one could address the family puzzle by taking different
fermion generations as bound to different nodes and fixing the
number of nodes in the bulk at three.
For the case of a decreasing warp factor, and infinite number of
anti-de-Sitter island-universes, one has something like the landscape picture
in string theory/M theory \cite{landscape, landscape1, landscape2}, with its large number of
universes with different parameters.

In the model presented in this paper ordinary matter
fields are assumed to be bound to one of the anti-de-Sitter `islands'. This is an
advantage over the original one-brane or two-brane models of
\cite{brane1, brane2, brane3, brane4}, where the ordinary matter fields are assumed to be
localized to a brane with positive or negative tension. In these
models one needs to explain why these brane tensions are not
observed.

There are several open questions or problems with the present 5D
standing wave solution as brane world model:
The first question concerns the stability of the whole 5D solution given the
presence of the ghost-like scalar field. This issues of the
stability of brane world models in the presence of sources which
violate some or all of the usual energy conditions (e.g. negative
tension branes \cite{brane1, brane2, brane3, brane4}, or ghost fields \cite{Koley, Koley1, Koley2, Phantom})
has been known since the beginning of the study of brane world
models \cite{pilo}. In section $3$ we present a possible resolution to this issue
by considering Weyl's generalization of Riemannian geometry. We also offer
some discussion in the conclusion section of cases where a
ghost field, at least at the classical level, leads to greater
stability of brane world models as compared to normal scalar fields.
The second question is: ``On or near the nodes are the gravitational
fluctuations confined so as to give effective 4D gravity near the
nodes?". In section $6$ will show that near the nodes one does get
effective 4D gravity in much the same way as in \cite{brane1, brane2, brane3, brane4}.
The last question/problem is give a localization mechanism for the
matter fields (i.e. fields with spin 0, spin $\frac{1}{2}$, spin 1) on the brane. This
localization problem has been an unresolved question for brane world
models in general \cite{Local, Local1, Local2, Local3, Local4, Local5, Local6}. Here we simply postulate that matter
fields are bound to the nodes of the standing wave solutions. In
section $7$ we give two possible mechanisms for accomplishing this
binding of matter fields to the nodes. The first mechanism involves
the coupling of the quadrupole moment of the stress-energy tensor of
the matter fields to the Riemann tensor. The second mechanism
involves coupling of matter fields to the ghost-like field.


\section{Metric and Einstein's equations}

We study solutions of the 5-dimensional Einstein equations,
\begin{equation} \label{5d-einstein}
R_{ab} - \frac{1}{2} g_{ab} R = \frac{8 \pi G_5}{c^4} T_{ab} -
\Lambda _5 g_{ab}~,
\end{equation}
where $G_5$ and $\Lambda _5$ are 5D Newton and cosmological
constants respectively \cite{brane1, brane2, brane3, brane4}. Lower case Latin indices $a, b$
run  over the full 5D space-time $0,1,2,3,4$. The metric {\it
ansatz} we use is
\begin{equation} \label{metric}
ds^2 = e^{2a|r|}\left( dt^2 - e^{u}dx^2 - e^{u}dy^2 - e^{-2u}dz^2
\right) -dr^2~,
\end{equation}
where $a$ is a constant and the {\it ansatz} function $u = u(t,r)$
only depends on the time $t$ and the extra coordinate $r$. This {\it
ansatz} is a combination of the 5D warped brane world model through the
$e^{2a|r|}$ term \cite{brane1, brane2, brane3, brane4} plus an anisotropic ($t,r$)-dependent warping of the brane coordinates,
$x,y,z$, through the terms $e^{u(t,r)}, e^{-2u(t,r)}$. The absolute value sign around $r$ gives a
$\delta$-function source/brane at $r=0$ exactly like the one brane
models of \cite{brane1, brane2, brane3, brane4}. Since we will focus on standing wave solutions `to
the right', i.e. $r>0$, we drop the absolute value sign in
\eqref{metric}. We study both decreasing ($a<0$) and increasing
($a>0$) warp factors. We also find that for both increasing and decreasing
warp factors the ansatz function $u(t,r)$ has nodes -- $u(t,r) =0$ -- 
at specific values of $r$. At these nodes the space-time in \eqref{metric} 
reduces to 4D Minkowski space-time plus a scaled, negative cosmological constant -- 
at the nodes the space-time becomes effectively 4D anti-de-Sitter.
As one moves away from these nodes there is an anisotropic stretching/shrinking 
along the $x,y,z$ directions due to the metric components $e^{u(t,r)}, e^{-2u(t,r)}$. 
If $u(t,r)$ is positive (negative) as one moves
away from the node in $r$ the $x,y$ directions will expand (shrink) as $e^u$ while the
$z$ direction will shrink (expand) as $e^{-2u}$. In section $7$ we will use this 
feature of the metric to propose a possible localization mechanism for matter fields.

For the {\it ansatz} \eqref{metric} we find the non-zero components
of the Ricci tensor:
\begin{eqnarray} \label{field-eqns2}
R_{tt} &=& e^{2ar} \left( -\frac 32 e^{-2ar} \dot u ^2 + 4a^2 \right)~, \nonumber \\
R_{xx}&=&R_{yy}=e^{2ar+u}\left( \frac 12 e^{-2ar}\ddot u - 2au'-\frac 12 u'' -4a^2\right)~, \nonumber \\
R_{zz} &=&e^{2ar-2u}\left( - e^{-2ar}\ddot u + 4au' + u'' - 4a^2\right)~, \\
R_{rr} &=& -\frac 32 u'^2 - 4a^2~, \nonumber\\
R_{rt} &=& -\frac 32 \dot uu' ~, \nonumber
\end{eqnarray}
where overdots mean time derivatives and primes stand for
derivatives with respect to $r$. In \eqref{field-eqns2} there should
be terms proportional to $\delta (r)$ coming from the tension of
the thin brane at $r=0$. We do not write this explicitly since we
focus on $r>0$ and have thus dropped the absolute value in the warp
factor of the metric \eqref{metric}.


\section{Source scalar field}

To the metric above we add a massless, non-self interacting scalar
phantom/ghost-like field, $\phi (t,z)$ \cite{phantom, phantom1}, which
obeys the Klein-Gordon equation,
\begin{equation} \label{phi}
\frac{1}{\sqrt{g}}~\partial_a (\sqrt{g} g^{a b}\partial_b \phi) =
e^{-2 a r}\ddot \phi - \phi'' - 4 a \phi' = 0 ~.
\end{equation}
Here $g$ is the determinant of the 5-dimensional background
space-time given by \eqref{metric}. The energy-momentum tensor of
the phantom-like field $\phi$ is taken in the form:
\begin{equation} \label{emt-phi}
T_{ab} = - \partial_a \phi\partial_b \phi + \frac 12 g_{a b} \partial^c
\phi~\partial_c \phi~.
\end{equation}
Strictly speaking $\phi$ is not a phantom field as defined in
\cite{caldwell}, where the criterion for a phantom field was $p/\rho
<-1$ ($p$ and $\rho$ are the pressure and energy density of the
field respectively). From \eqref{emt-phi} one can obtain $p$ and
$\rho$ for the field $\phi$ and since the field is non-self
interacting one does not have $p/\rho <-1$.

To avoid the well-known problems with stability that occur with
ghost fields we can associate the ghost-like field $\phi$ with the geometrical scalar field in
a five-dimensional, integrable Weyl model. In Weyl's model a massless scalar appears
through the definition of the covariant derivative of the metric
tensor,
\begin{equation} \label{D}
D_{c} g_{ab} = g_{ab}\partial_c \phi ~.
\end{equation}
This is a generalization of the Riemannian case where the covariant derivative
of the metric is zero. The result in \eqref{D} implies that 
the length of a vector is altered by parallel
transport. Weyl's scalar field in \eqref{D} may imitate a massless scalar field -- either 
an ordinary scalar or ghost-like scalar \cite{Weyl, Weyl1, Weyl2, Weyl3}. The gravitational
action for Weyl's 5D integrable model can be written as
\begin{equation}\label{grav-weyl}
S_g = \frac{1}{16 \pi G_5} \int d^5x \sqrt{g}\left[R -
(6-5\xi)~\partial^a \phi~\partial_a \phi\right]~,
\end{equation}
where $\xi$ is an arbitrary constant which can take any
value. For example, if one takes $\xi =13/10$ in \eqref{grav-weyl}
then the coefficient in front of $\partial^a \phi~\partial_a \phi$ becomes
$-1/2$. This would give a ghost field which would lead exactly 
to the equation of motion \eqref{phi} and the energy-momentum tensor
\eqref{emt-phi} for our phantom-like scalar field. Thus we can start
with a 5D Weyl model and require that we have Riemann geometry on the
brane by assuming that the geometrical scalar $\phi$ is independent
of brane coordinates $x^i$ and vanishes on the brane. We will show later that
our solution has exactly this character -- the scalar field $\phi$ vanishes
at the nodes of $u (t,r)$ and is independent of $x,y,z$. Associating 
our scalar field with the geometrical Weyl scalar defined via \eqref{D} avoids 
the usual instability problems of ghost fields since the Weyl model is known to be 
stable for any value of $\xi$.


\section{The solution}

For \eqref{emt-phi} one can rewrite the Einstein equations
(\ref{5d-einstein}) in the form:
\begin{equation}
\label{field-eqns1} R_{ab} = - \partial_a \sigma\partial_b \sigma
+\frac{3}{2} g_{ab} \Lambda _5~,
\end{equation}
where the gravitational constant has been absorbed via the
redefinition of the scalar field,
\begin{equation}
\label{sigma}
\sigma =\frac{\sqrt{8 \pi G_5}}{c^2}\phi ~.
\end{equation}

By combining \eqref{field-eqns1} and \eqref{field-eqns2} one can see
that the ``constant" terms from $R_{\mu \nu}$ (i.e. those terms that
up to some metric factor are $\pm 4 a^2$) can be canceled if the 5D
cosmological constant is chosen as
\begin{equation}
\label{L=a} \Lambda_ 5 =  \frac{8}{3} a^2~.
\end{equation}
This is the same as the fine tuning used in standard 5D brane models
\cite{brane1, brane2, brane3, brane4}.

The terms from $R_{\mu \nu}$ which are quadratic in $u$ can be
accounted for if one assumes that
\begin{equation} \label{sigma=u}
\sigma (t,r) = \sqrt{\frac{3}{2}}~u(t,r)~,
\end{equation}
so that $u(t,r)$ satisfies \eqref{phi}. A similar equality between
the scalar field and metric ansatz function was found for the domain
wall plus standing wave solutions of \cite{domain-wall}. Since our scalar field
is proportional to the metric ansatz function $u(t,r)$ the scalar field
will vanish wherever $u(t,r)$ has nodes. This requirement, that the scalar field
vanish at the zeros of $u(t,r)$, was one of the necessary conditions pointed out in the last
section in order to be able to associate our scalar field with a Weyl scalar field. 

Because of \eqref{L=a} and \eqref{sigma=u}, solving
Einstein equations has been reduced to finding solutions to the
ordinary differential equation,
\begin{equation} \label{field-eqns3}
e^{-2ar}~\ddot u - u'' - 4au' = 0~.
\end{equation}

We want to have a standing wave solution to \eqref{field-eqns3}, so
we use separation of variables writing
\begin{equation} \label{separation}
u(t,r) = C  \sin (\omega t) f(r)~,
\end{equation}
where $C$ and $\omega$ are constants. Because of
(\ref{sigma=u}) the same separation applies to $\sigma$, but with a
different constant, $C \rightarrow \sqrt{\frac{3}{2}} C$. Equation \eqref{field-eqns3} 
now becomes
(\ref{separation}) is:
\begin{equation}
\label{f}
f'' + 4 a f' + \omega^2 e^{-2 a r} f = 0 ~.
\end{equation}
The general solution to this equation is:
\begin{equation} \label{solution}
f(r) = A e^{-2ar} J_2 \left( \frac{\omega}{a} e^{-ar} \right) +
B  e^{-2ar} N_2 \left( \frac{\omega}{a} e^{-ar} \right) ~,
\end{equation}
where $A,B$ are constants and $J_2$ and $N_2$ are 2$^{nd}$ order
Bessel functions of the first and second kind respectively. Normally
the $N_2$ Bessel functions are discarded since they blow up at the
origin. But here the functional dependence is $e^{-a r}$, rather
than $r$, and neither $J_2$ nor $N_2$ diverges at $r=0$.

Before moving on to the discussion of the physical meaning of the solutions
in \eqref{solution} we add the boundary condition that the ghost-like
field should vanish at the brane, $r=0$. To accomplish this we
should take either $A=0$ or $B=0$, since the zeros of $J_2$ and
$N_2$ do not coincide. Then we set
\begin{equation} \label{omega}
\frac{\omega}{a} = X_{2,n}~,
\end{equation}
where $X_{2,n}$ is the $n^{th}$ zero of the 2$^{nd}$ order Bessel
function $J_2$ or $N_2$, depending on whether one takes $A=0$ or
$B=0$ in \eqref{solution}. The condition \eqref{omega} quantizes the
oscillation frequency, $\omega$.


\section{Standing waves as Anti-de-Sitter `islands'}

We now analyze the physical consequences for metric
\eqref{metric} given the solutions given by \eqref{separation},
and \eqref{solution}.

First we note that \eqref{metric} in terms of the behavior of the
$x, y, z$ coordinates gives different asymptotic properties for the
increasing ($a>0$) and decreasing ($a<0$) warp factor cases. For the
increasing warp factor ($a>0$) the metric function, $u(t,r) \propto
f(r)$, decreases like $e^{-2ar}$ as one moves into the bulk. As one
goes to large $r \rightarrow \infty$ this factor of $e^{-2ar}$
drives $u(t,r) \rightarrow 0$ and one has an anti-de-Sitter space-time but
scaled up by an overall factor $e^{2 a r}$. For the decreasing
factor ($a<0$) the function $u(t,r)$ grows like $e^{2ar}$ into the
bulk therefore distances in the $x$ and $y$ directions expand like
Exp$(e^{2ar})$, while in the $z$ direction they shrink like Exp$(-2
e^{2ar})$.

The second observation is that $u(t,r)$ is oscillatory and has zeros at
various points along the $r$-direction -- whenever $f(r)=0$. This
happens when $J_2 (r) =0$ or $N_2 (r) =0$, depending on if one is
considering $B=0$ or $A=0$. At these zero points, $r_m$, one has $u(t, r_m) =0$ 
and the metric in \eqref{metric} reduces to the standard 5D brane
metric with an exponential warp factor \cite{brane1, brane2, brane3, brane4}. This spatial
oscillatory behavior of $u (t,r)$ leads to `islands' of
anti-de-Sitter space-times in the bulk.

The third observation is that the physical parameters on each 
node are scaled by the warp factor $e^{ar}$ or some power thereof.
For example, taking the 4D reduction of 5D Einstein equations \eqref{5d-einstein}
one finds that on the nodes of the anti-de-Sitter `islands' there are effective
4D, negative cosmological constants given by
\begin{equation}
\label{cc} \Lambda _4 = e^{2 a r_m} \Lambda _5 ~.
\end{equation}
Thus the values of the effective 4D cosmological constant on the nodes
is set by the scaled 5D cosmological constant. In the usual thin brane
models \cite{brane1, brane2, brane3, brane4} the 5D cosmological constant is fine tuned
to exactly cancel the 4D brane tension. Thus on the brane the
effective 4D cosmological constant is zero. In our case at the
nodes, $r_m$, there is no brane tension and so our effective
4D cosmological constant is non-zero and given by \eqref{cc}. For a decreasing
wrap factor (i.e. $a<0$) one has a cosmological constant which is
exponentially suppressed relative to the true 5D value, $\Lambda_5$. 
Below we argue that these nodes are 4D anti-de-Sitter island
universes on which matter fields are bound and on which gravity is
effectively 4D.

For the case $a>0$ and $B=0$, there will be $(n+1)$ such `islands' ($n$
is the number of the zero from \eqref{omega}). As $r$ runs from $0$
to $\infty$ the argument of $J_2$ in \eqref{solution} runs from
$X_{2, n}$ to $0$ giving $n$ zeros. One additional zero comes from
$f(r \rightarrow \infty) =0$. The zeros will occur at $r_m$ where $\omega e^{-a
r_m} / a = X_{2, m}$ with $X_{2, m}$ being a zero of $J_2$ with
$m<n$. The case $a>0$ and $A=0$ works out the same
with $(n+1)$ `islands' since one also has an additional zero coming
from $f(r \rightarrow \infty) \rightarrow 0$ -- in this case the divergence 
coming from $N_2 (0)$ is dominated by the $e^{-2ar}$ pre-factor in \eqref{solution}.
The $\omega e^{-ar}/a $  dependence of $f(r)$ has
the effect of stretching out the zeros of the Bessel functions, i.e
the spatial oscillation frequency decreases with $r$.

In the case $a<0$ (with either $A=0$ or $B=0$) there are an infinite number of zeros for $u(t,r)$
due to the $\omega e^{ar}/a $ dependence of $f(r)$. In addition the
zeros are compressed as $r$ increases, i.e. the spatial oscillation
frequency increases with $r$. This compression ($a>0$) and stretching ($a<0$) of the location of the nodes
has a connection with the effective thickness of the 4D anti-de-Sitter islands -- the compression of the nodes
tends to lead to `islands' with a smaller effective thickness while stretching of the nodes leads to
`islands' with a large effective thickness. We discuss this further in section $7$ where we examine
localization mechanisms for matter fields to the nodes. 


\section{Localization of gravity}

Near $r=0$ in metric \eqref{metric} one has a $node + brane$
as in the 5D single brane models of
\cite{brane1, brane2, brane3, brane4}. The gravitational perturbations around $r=0$ will have
one delta-function bound state and continuum states which start from
zero mass i.e. which do not have a mass gap. Thus at the $r=0$ node
one has  effective 4D gravity as in the
original models. We now look at the nodes of $u(r,t)$ for $r
> 0$ to see if near these nodes gravity also becomes approximately 4D.
This is done by studying the 4D perturbations of the metric near
the nodes. If there is a zero mass graviton mode near the node then
its exchange between particles localized on the node will lead to a
Newtonian potential with corrections coming from the massive modes ($m>0$).
Close to the nodes one has $u(r,t) \approx 0$ and the metric
\eqref{metric} takes on the usual 5D warped geometry.
Near the nodes small fluctuations around this
background can be written as
\begin{equation}
\label{node}
ds^2 \approx \left[ e^{2 a r} \eta_{\mu \nu} + h_{\mu \nu} (x_\mu , r) \right] dx^\mu dx^\nu - dr^2 ~,
\end{equation}
where $x_\mu = (t,x,y,z)$ and Greek indices run over 4D space-time --
$\mu , \nu = 0,1,2,3$. The tensor, $h_{\mu \nu}$, gives 4D
perturbations around the usual 5D brane world background.
Further we fix the gauge so that $h_{\mu \nu}$ is
transverse and traceless,
\begin{equation}
\partial _\mu h^{\mu} _\nu = h^\mu _\mu =0 ~.
\end{equation}
Next we separate the tensor perturbation into a 1D and 4D part as
\begin{equation}
h_{\mu \nu} (x^\mu, r) = \psi (r) h ^{(4)} _{\mu \nu} (x^\mu )~,
\end{equation}
with $h ^{(4)} _{\mu \nu} (x^\mu )$ satisfying
\begin{equation}
\Box ^{(4)} h ^{(4)} _{\mu \nu} (x^\mu ) = p^2= m^2~.
\end{equation}
Putting all this together yields the following equation for the 1D
part of the perturbations
\begin{equation}
\label{pert-1D}
\psi '' - 4 a^2 \psi - m^2 e^{-2ar} \psi =0 ~.
\end{equation}
In order for gravity to be effectively 4D near the nodes
\eqref{pert-1D} should have a zero mode solution -- $m=0$. It is easy
to see that it does have a zero mode given by,
\begin{equation}
\psi _0 (r) = e^{2ar}~,
\end{equation}
up to a normalization constant. Equation \eqref{pert-1D} also has $m
\ne 0$ solutions which are similar to \eqref{solution} i.e. combinations
of second order Bessel functions of the first and second kind, $J_2
, N_2$, \cite{Rub}. The exchange of the zero mode, $m=0$, between
massive particles fixed on the node, will lead to a $1/r$ Newtonian
potential while the exchange of the $m \ne 0$ modes will lead to
corrections which go as $1/(a^2 r^2)$ \cite{brane1, brane2, brane3, brane4, Rub}.

The effective 4D Newton's constant on a particular node can be
obtained using the above results. Let us focus on one particular
node, $r_m$. The massless gravitational perturbation is
\begin{equation}
\label{h-shift}
h_{\mu \nu} = e ^{2 a r} h ^{(4)} _{\mu \nu} (x^\mu ) ~,
\end{equation}
where $h ^{(4)} _{\mu \nu} (x^\mu )$ is the 4D part. The effective
4D Newton's constant can be obtained by inserting \eqref{h-shift}
into the 5D gravitational action and looking at the quadratic part
\begin{eqnarray}
\label{grav-action}
S_g &=& \frac{1}{16 \pi G_5} \int _ {r_m - d} ^{r_m +d} \frac{dr}{e^{2 a r}} \int d^4x \left( \partial h \right)^2 \nonumber \\
&=& \frac{1}{16 \pi G_5} \int _{r_m - d} ^{r_m +d} dr ~ e^{2 a r } \int d^4 x \left( \partial h ^{(4)} \right)^2 \\
&=& \frac{e^{2ar_m} \sinh (2 a d)}{16 \pi G_5 a} \int d^4 x \left( \partial h ^{(4)} \right)^2 ~, \nonumber
\end{eqnarray}
where $2d$ is the width around the node, $r_m$, within which the
matter particles are assumed to be bound. The index $m$ satisfies $m
\le n$ where for $m=n$ one is located at the $r=0$ brane.
Using the last expression in \eqref{grav-action} the effective 4D
Newton's constant can be written as
\begin{equation}
\label{4D-newton}
G_4 = \frac{a G_5}{e^{2ar_m} \sinh (2 a d)}  \approx \frac{G_5}{2 ~ d ~e^{2 a r_m}}~.
\end{equation}
The last expression assumes a small thickness i.e. $d \ll 1$, or $a
d \ll 1$. In the case when $a>0$ one can exponentially suppress the
4D Newton's constant, $G_4$,  relative to the 5D Newton's constant,
$G_5$.


\section{Localization of matter}

So far in this paper we have assumed some mechanism for binding matter fields
to the nodes of the standing waves. Now we want to give two possible
localization mechanisms.

The first mechanism is borrowed from condensed matter physics. It is
known that standing electromagnetic waves, so called optical
lattices, can provide trapping of various particles by
scattering and dipole forces \cite{Opt, Opt1}. In \cite{quadr, quadr1}
localization was demonstrated through quadrupole forces as well. It
is also known that the motion of test particles in the field of a
gravitational wave is similar to the motion of charged particles in
the field of an electromagnetic wave \cite{Ba-Gr}. Thus standing
gravitational waves could also provide confinement of matter via
quadrupole forces. As an example let us consider the equations of
motion of the system of spinless particles in the quadrupole
approximation \cite{Dix, Dix1, Dix2},
\begin{equation}
\label{quad}
\frac{Dp^\mu}{ds}= F^\mu_{quad} = -\frac 16
J^{\alpha\beta\gamma\delta}D^\mu R_{\alpha\beta\gamma\delta}~,
\end{equation}
where $p^\mu$ is the total momentum of the matter field/particle and
$J^{\alpha\beta\gamma\delta}$ is the quadrupole moment of the
stress-energy tensor for the matter field/particle. 
The oscillating metric due to gravitational
waves should induce a quadrupole moment in the matter fields. If the
induced quadrupole moment is out of phase with the gravitational
wave the system energy increases in comparison with the resonant
case and the fields/particles will feel a quadrupole force,
$F^\mu_{quad}$, which ejects them out of the high curvature region i.e.
it would localize them at the nodes. This
gravitational binding mechanism would be effective for all types of
matter fields since gravity couples to all forms of energy-momentum.
Since the matter fields are ejected from the high curvature region and toward the
nodes the width of the anti-de-Sitter `islands' (i.e. the thickness of the branes)
depends on how rapidly the ansatz function $u(t,r) \propto f(r)$ from
\eqref{separation} \eqref{solution} changes from zero. The width of some
anti-de-Sitter `island' will depend in the derivative of $f(r)$ at the node --
$f'(r_m)$. The larger $f'(r_m)$ is the smaller the width of the anti-de-Sitter `island'.
Now $f'(r_m)$ depends on the size of either $A$ or $B$ and  as well as 
$a$, $\omega$ and $r_m$. For some choice of these parameters some nodes might not
be phenomenologically acceptable since the brane thickness might be too large.
Also in general for the case when $a>0$ there will be fewer phenomenologically
acceptable branes since in this situation the nodes are stretched out, as discussed in
section $5$, and this tends to decrease $f'(r_m)$ as $r$ increases. On the other hand 
when $a<0$ there is an increase of $f'(r_m)$ as $r$ increases which leads to more 
nodes which have a phenomenologically acceptable thickness.

The second mechanism would be to propose some coupling between the
ghost-like field, $\sigma$, and matter fields of the form $\sigma A
A$, where $A$ is a scalar, spinor, vector or tensor field. For a
normal scalar field such a coupling leads to an attractive force,
while for a ghost-like field it leads to repulsion of the matter
fields from regions with non-zero $\sigma$. This would force the
fields $A$ to congregate at regions were $\sigma$ vanishes, i.e. the
nodes of the standing waves. Note that coupling ordinary matter
fields to ghost fields is problematic since in general it leads to
instabilities. However, as mentioned in the Sec. $3$, our
massless ghost-like field can be associated with the geometrical
scalar of integrable Weyl models.

Both binding mechanisms have open questions (e.g.``Does the
gravitational standing wave induce a quadrupole moment of the matter
field stress-energy tensor and if so what is its form?"; ``Can one
safely couple ordinary matter fields to ghost-like fields?"). A
detail analyzes of localization of specific matter fields via these
two possible mechanisms will be considered in a forthcoming paper.


\section{Conclusions and discussion}

A simple standing wave solution for 5D spacetime with a ghost-like
scalar field plus 5D negative cosmological constant was presented.
The requirement that the ghost-like field vanish on the brane
quantized the standing wave oscillation frequency in terms of the
zeros of the 2$^{nd}$ order Bessel functions. Thus the ghost-like
field is not observable on the brane, at $r=0$, nor at any of the
nodes, $r_m$, of the standing wave. This is similar to the static brane
model \cite{Koley, Koley1, Koley2}, where the phantom/ghost field only existed in
the bulk.

For the case of increasing (decreasing) warp factors the ghost-like
field, $\sigma (t,r)$, and the metric function, $u(t,r)$, vanish at
a finite (infinite) number of places in the bulk forming `islands'
of 4D anti-de-Sitter space-time. By assuming that the matter fields are bound
to these nodes we arrive at a model where each node is an anti-de-Sitter
island-universe with different parameters, i.e. an effective 4D
Newton's constant from \eqref{4D-newton} and an effective 4D
cosmological constant from \eqref{cc} scaled by the value $e^{2 a
r_m}$ at the particular node. One could address the hierarchy
problem and the small size of the cosmological constant by taking
our universe as one of these Bessel nodes, where the 5D Newton and
cosmological constants are appropriately scaled to their effective
4D values. A problem is that the scalings go in oppose directions.
For example, with $a>0$ \eqref{cc} gives a 4D cosmological constant
which is exponentially larger than the 5D cosmological constant,
while \eqref{4D-newton} gives a 4D Newton's constant which is
exponentially smaller than the 5D Newton's constant. For $a<0$ the effective
4D Newton's constant and the effective 4D cosmological constant have opposite 
scaling to the $a>0$ case.

There are two distinctly different cases for the standing wave
solutions  -- increasing warp factor ($a>0$) and decreasing warp
factor ($a<0$) -- corresponding to a finite, or infinite number of anti-de-Sitter
`islands' respectively.  The finite case could be applied to the
generation problem by fixing the model to have three `islands' with
different fermion generations bound to different nodes. This is
different from brane world models of the generation puzzle like
\cite{gog-sing, gog-sing1, gog-sing2}, where the generations were associated with
different zero modes bound to a single brane. For the infinite
branes case one has a simple version of the landscape picture coming
from string/M theory, where one has a large number of anti-de-Sitter
island-universes each having different parameters e.g. effective 4D
gravitational and 4D cosmological constants.

The present models has the same kind of warped geometry as the usual
brane models \cite{brane1, brane2, brane3, brane4}. In distinction from these models the
``branes" (i.e. the nodes of $u(r,t)$, excluding the one at $r=0$) of
the present standing wave background solution do not have a brane
tension but are simply 4D anti-de-Sitter space-times. This is an advantage
since for the models of \cite{brane1, brane2, brane3, brane4} one should explain why the
$\delta$-source brane tension is not observed. The disadvantage of
the present model is the need for a ghost-like field. The need for
the ghost-like field is ameliorated by the fact that it vanishes on
the $r=0$ brane and on all the Bessel node anti-de-Sitter `islands'.

The issue of instability due to an unusual matter source is also a
problem for the original two-brane model which has a negative
tension brane. If the negative tension brane is free to oscillate
this results in arbitrarily large negative energy modes making the
system unstable \cite{pilo}. A similar problem occurs in the
present model if we take our scalar field as an ordinary ghost field. 
However, if we associate our ghost field with the geometrical
scalar of a 5D integrable Weyl model \cite{Weyl, Weyl1, Weyl2, Weyl3} this alleviates some of the
instability problems since the scalar field comes from the
metric via \eqref{D}. In addition for Weyl models the
ghost-like field does not have the standard couplings with the brane
energy-momentum and thus does not cause instabilities.

Additionally previous work with standard ghost fields shows -- that at least
at the classical level -- bulk ghost fields may in fact be better at
stabilizing brane world models as compared to bulk regular scalar
fields. In \cite{pospelov} bulk ghost fields in 5D are shown to lead
to radion stabilization for models with positive tension branes.
Essentially the bulk ghost field replaces the negative tension
brane. Further it was shown that a bulk scalar field whose sign can
vary between regular (positive) and ghost (negative) can be used to
model both positive and negative tension branes. Also, there are
certain configurations of the bulk ghost fields of reference
\cite{pospelov} which lead to stronger localization of gravity than
the original brane models. In references \cite{maity, maity1} it is shown
that the two brane model of \cite{brane1, brane2, brane3, brane4} is stabilized by a
ghost/tachyon bulk field, while for a regular bulk scalar field the
two brane model is unstable. While the stability issue is crucial
for all brane world models it has not yet found a complete solution
even for the original models of \cite{brane1, brane2, brane3, brane4}. Here, as in
\cite{Koley, Koley1, Koley2, pospelov} we simply use the ghost scalar to build a
brane world model from a 5D standing wave, leaving the generally
unresolved question of stability of the solution for a future work.
As a final note in \cite{sen, sen1, sen2} it is shown how to construct a
consistent field theory with tachyons in the context of D-branes.

Although throughout this paper we have assumed some mechanism for binding 
matter fields/particles to the anti-de-Sitter `islands' in section $7$ we
put forward two possible mechanisms for accomplishing this binding.
The first localization mechanism was based on the quadrupole force in \eqref{quad} 
which would eject particles from the high curvature regions and thus drive them
to the nodes. One might give a heuristic description of this binding mechanism 
by noting that as one moves away from the nodes where $u(t,r)=0$ that the
brane coordinates $x,y,z$ are distorted in an anisotropic way -- the $x,y$ coordinates
will be stretched/compressed while the $z$ coordinate will be compressed/stretched.
This anisotropy causes a force, via \eqref{quad}, which tends to drive the matter
fields toward the nodes. The second mechanism also involved ejecting the matter fields
from the anti-nodes and towards the nodes, but in this case the ejection mechanism was
accomplished by coupling the matter fields to the ghost-like scalar. This would give
rise to a repulsive force which would drive the matter fields toward regions where the
ghost-like scalar field was small i.e. toward the nodes. For both mechanisms the
thickness of the brane was related to how rapidly the metric and scalar field ansatz
functions changed near the nodes i.e. to have a thinner brane one should make $f'(r_m)$
larger. In general the case with $a<0$ would have thinner branes.

\medskip


\noindent {\bf Acknowledgments:}

M. G. and D. S. are supported by a 2008-2009 Fulbright Scholars
Grants. M. G. acknowledges the grant of Georgian National Science
Foundation $ST~09.798.4-100$. D. S. would like to thank Vitaly Melnikov for the
invitation to work at the Institute of Gravitation and Cosmology at
Peoples' Friendship University of Russia.



\begin{thebibliography}{99}

\bibitem{brane1} 
M. Gogberashvili,Int. J. Mod. Phys. {\bf D 11} (2002) 1635,
arXiv: hep-ph/9812296.

\bibitem{brane2} 
M. Gogberashvili, Mod. Phys. Lett. {\bf A 14} (1999) 2025,
arXiv: hep-ph/9904383.
   
\bibitem{brane3}             
L. Randall and R. Sundrum, Phys. Rev. Letts. {\bf 83} (1999) 3370,
arXiv: hep-ph/9905221.
  
\bibitem{brane4}               
L. Randall and R. Sundrum, Phys. Rev. Letts. {\bf 83} (1999) 4690,
arXiv: hep-th/9906064.

\bibitem{Rub} 
V.A. Rubakov,Phys. Usp. {\bf 44} (2001) 871,
arXiv: hep-ph/0104152.

\bibitem{GoSi} 
M. Gogberashvili and D. Singleton, Phys. Lett. {\bf B 582} (2004) 95,
arXiv: hep-th/0310048.
 
 \bibitem{GoSi1}
M. Gogberashvili and D. Singleton, Phys. Rev. {\bf D 69} (2004) 026004,
arXiv: hep-th/0305241.

\bibitem{Bronn}
S. Abdyrakhmanov, K. Bronnikov and B.E. Meierovich, Grav. Cosmol. {\bf 11} (2005) 82,
arXiv: gr-qc/0503055.

\bibitem{dzh-sing}                
V. Dzhunushaliev, V. Folomeev, D. Singleton and S. Aguilar-Rudametkin,
Phys. Rev. {\bf D 77} (2008) 044006, arXiv: hep-th/0703043.

\bibitem{Koley} R. Koley and S. Kar, Mod. Phys. Lett. {\bf A 20} (2005) 363,
arXiv: hep-th/0407159.

\bibitem{Koley1}               
R. Koley and S. Kar, Phys. Lett. {\bf B 623} (2005) 244,
arXiv: hep-th/0507277.
                
\bibitem{Koley2} 
R. Koley and S. Kar, Class. Quant. Grav. {\bf 24} (2007) 79,
arXiv: hep-th/0611074.

\bibitem{Phantom} 
M. Szydlowski, W. Czaja and A. Krawiec, Phys. Rev. {\bf E 72} (2005) 036221,
arXiv: astro-ph/0401293.
                  
\bibitem{Phantom1}                 
E. Elizalde, S. Nojiri and S.D. Odintsov, Phys. Rev. {\bf D 70} (2004) 043539,
arXiv: hep-th/0405034.
                  
\bibitem{Phantom2}
V. Dzhunushaliev, V. Folomeev, K. Myrzakulov and R. Myrzakulov,
Int. J. Mod. Phys. {\bf D 17} (2008) 2351,
arXiv: gr-qc/0608025.
                  
\bibitem{S} 
M. Gutperle and A. Strominger, JHEP {\bf 0204} (2002) 018,
arXiv: hep-th/0202210.

\bibitem{S1}          
C.M. Chen, D.M. Gal'tsov and M. Gutperle, Phys. Rev.  {\bf D 66} (2002) 024043,
arXiv: hep-th/0204071.
            
\bibitem{S2}
M. Kruczenski, R.C. Myers and A.W. Peet, JHEP {\bf 0205} (2002) 039,
arXiv: hep-th/0204144.
            
\bibitem{S3}
N.S. Deger and  A. Kaya, JHEP {\bf 0207} (2002) 038,
arXiv: hep-th/0206057.

\bibitem{S4}
V.D. Ivashchuk and D. Singleton, JHEP {\bf 0410} (2004) 061,
arXiv: hep-th/0407224.

\bibitem{Weyl} 
J.M. Salim and S.L. Sauti, Class. Quant. Grav. {\bf 13} (1996) 353.

\bibitem{Weyl1}
M. Israelit, Found. Phys. {\bf 35} (2005) 1725,
arXiv: 0710.3690v1 [gr-qc].

\bibitem{Weyl2}
J.E. Madriz Aguilar and C. Romero, Found. Phys. {\bf 39} (2009) 1205,
arXiv: 0809.2547 [math-ph].

\bibitem{Weyl3}
J.E. Madriz Aguilar and C. Romero, Int. J. Mod. Phys. {\bf A 24} (2009) 1505,
arXiv: 0906.1613 [gr-qc].

\bibitem{landscape} 
M. Douglas, JHEP {\bf 0305} (2003) 046, arXiv: hep-th/0303194.
   
\bibitem{landscape1}                 
S. Ashok and M. Douglas,
JHEP {\bf 0401} (2004) 060, arXiv: hep-th/0307049.

\bibitem{landscape2}
L. Susskind, in {\it Universe or multiverse?} B. Carr (ed.)
(Cambridge Univ. Press, Cambridge 2007), arXiv: hep-th/0302219.

\bibitem{pilo} 
L. Pilo, R. Rattazzi and A. Zaffaroni, JHEP {\bf 0007} (2000) 056,
arXiv: hep-th/0004028.

\bibitem{Local}
B. Bajc and G. Gabadadze, Phys. Lett. {\bf B 474} (2000) 282; arXiv: hep-th/9912232.
 
\bibitem{Local1}              
A. Pomarol, Phys. Lett. {\bf B 486} (2000) 153; arXiv: hep-ph/9911294.

\bibitem{Local2}
I. Oda, Phys. Rev. {\bf D 62} (2000) 126009, arXiv: hep-th/0008012.

\bibitem{Local3}
I.Oda, Phys. Lett. {\bf B 571} (2003) 235, arXiv: hep-th/0307119.

\bibitem{Local4}               
M. Gogberashvili and P. Midodashvili, Phys. Lett. {\bf B 515} (2001) 447, arXiv: hep-ph/0005298.

\bibitem{Local5}
M. Gogberashvili and P. Midodashvili, Europhys. Lett. {\bf 61} (2003) 308, arXiv: hep-th/0111132.

\bibitem{Local6}
P. Kanti, R. Madden and K.A. Olive,
Phys. Rev. {\bf D 64} (2001) 044021, arXiv: hep-th/0104177.

\bibitem{phantom} 
K. Bronnikov, Acta. Phys. Pol. {\bf B 4} (1973) 251.

\bibitem{phantom1} 
H. Ellis, J. Math. Phys. {\bf 14} (1973) 104.

\bibitem{caldwell} 
R. R. Caldwell, Phys. Lett., {\bf B 545} (2002) 23,
arXiv: astro-ph/9908168.

\bibitem{domain-wall} 
M. Gogberashvili, S. Myrzakul and D. Singleton, Phys. Rev. {\bf D 80} (2009) 024040,
arXiv: 0904.1851 [gr-qc].

\bibitem{Opt} 
W. D. Phillips, Rev. Mod. Phys. {\bf 70} (1998) 721.

\bibitem{Opt1}               
H.J. Metcalf, and P. van der Straten, {\it Laser Cooling and Trapping}
(Springer, New York 1999).

\bibitem{quadr} 
N. Moiseyev, M. \v{S}indelka and L.S. Cederbaum, Phys. Lett. {\bf A 362} (2007) 215.

\bibitem{quadr1}
N. Moiseyev, M. \v{S}indelka and L.S. Cederbaum, Phys. Rev. {\bf A 74} (2006) 053420.

\bibitem{Ba-Gr} 
D. Baskaran and L.P. Grishchuk, Class. Quant. Grav. {\bf 21} (2004) 4041,
arXiv: gr-qc/0309058.

\bibitem{Dix} 
W.G. Dixon, Nuovo Cim. {\bf 34} (1964) 318.

\bibitem{Dix1}
W.G. Dixon, Proc. R. Soc. London {\bf A 314} (1970) 499.

\bibitem{Dix2}
W.G. Dixon, Gen. Rel. Grav. {\bf 4} (1973) 199.

\bibitem{gog-sing} 
N. Arkani-Hamed and M. Schmaltz, Phys. Rev. {\bf D 61} (2000) 033005,
arXiv: hep-ph/9903417.

\bibitem{gog-sing1}                  
S. Aguilar and D. Singleton, Phys. Rev. {\bf D 73} (2006) 085007,
arXiv: hep-th/0602218.
                  
\bibitem{gog-sing2} 
M. Gogberashvili, P. Midodashvili and D. Singleton, JHEP {\bf 0708} (2007) 033,
 arXiv: 0706.0676 [hep-th].

\bibitem{pospelov} 
M. Pospelov, Int. J. Mod. Phys. {\bf A 23} (2008) 881,
arXiv: hep-ph/0412280.

\bibitem{maity} 
D. Maity, S. SenGupta and S. Sur, Phys. Lett. {\bf B 643} (2006) 348,
arXiv: hep-th/0604195.

\bibitem{maity1}
A. Das, S. Kar and S. SenGupta, Int. J. Mod. Phys. {\bf A 24} (2009) 4457,
arXiv: 0804.1757 [hep-th].

\bibitem{sen} 
A. Sen, JHEP {\bf 0204} (2002) 048, arXiv: hep-th/0203211.

\bibitem{sen1}
A. Sen, JHEP {\bf 0207} (2002) 065, arXiv: hep-th/0203265.

\bibitem{sen2}
A. Sen, Mod. Phys. Lett. {\bf A 17} (2002) 1797, arXiv: hep-th/0204143.

\end{thebibliography}
\end{document}